\documentclass[12pt]{article}
\usepackage{geometry} 
\usepackage {ifthen,color,xypic,amssymb}
\usepackage{graphicx}
\usepackage{amsmath}
\usepackage{latexsym}
\def\hybrid{\topmargin -20pt  \oddsidemargin 0pt
                        \headheight 0pt   \headsep 0pt
                        \textwidth 6.25in 
                        \textheight 9.5in 
                        \marginparwidth .875in
                        \parskip 5pt plus 1pt   \jot = 1.5ex}

\hybrid

\def\x{\times}
\def\ox{\otimes}
\def\o+{\oplus}
\def\ra{\rightarrow}
\def\lra{\longrightarrow}
\def\Lra{\Longrightarrow}

\def\da{\downarrow}
\def\beqa{\begin{eqnarray}}
\def\eeqa{\end{eqnarray}}

\newcommand{\ov}{\overline}

\newcommand{\al}{\alpha}

\newcommand{\ga}{\gamma}

\newcommand{\De}{\Delta}
\newcommand{\la}{\lambda}
\newcommand{\si}{\sigma}
\newcommand{\ep}{\epsilon}

\newcommand{\G}{{\cal G}}
\newcommand{\ccL}{{\cal L}}
\newcommand{\cO}{{\cal O}}

\newcommand{\cV}{{\cal V}}

\def\Si{\Sigma}

\newcommand\iso{\kern.35em{\raise3pt\hbox{$\sim$}\kern -1.1em\to} \kern.3em}

\newcommand\fp{\times_{ B}}
\newcommand\rest[2]{{#1}_{\vert #2}}

\newcommand\Oc{{\mathcal O}}
\newcommand\Pc{{\mathcal P}}

\newcommand\V{{\mathcal V}}

\newcommand\X{{\widehat X}}

\newcommand\FM{Fourier-Mukai transform}

\DeclareMathOperator{\Tr}{Tr}
\DeclareMathOperator{\tr}{tr}

\DeclareMathOperator{\ch}{ch}

\newcommand{\resetcounter}{\setcounter{equation}{0}}

\begin{document}
\thispagestyle{empty}
\rightline{ASC-LMU 08/06}
\rightline{hep-th/0602247}
\vspace{2truecm}
\centerline{\LARGE Invariant Bundles on $B$-fibered Calabi-Yau Spaces}
\vspace{.4truecm}
\centerline{\LARGE and the Standard Model}

\vspace{1.5truecm}
\centerline{Bj\"orn Andreas$^1$ and Gottfried Curio$^2$}

\vspace{.6truecm}
\centerline{$^1${\em Institut f\"ur Mathematik und Informatik, 
Freie Universit\"at Berlin}}
\centerline{\em Arnimallee 14, 14195 Berlin, Germany}
\centerline{$^2${\em Arnold-Sommerfeld-Center for Theoretical Physics}}
\centerline{{\em Department f\"ur Physik, 
Ludwig-Maximilians-Universit\"at M\"unchen}}
\centerline{{\em Theresienstr. 37, 80333 M\"unchen, Germany}}

\vspace{1.0truecm}

\begin{abstract}
We derive the Standard model gauge group together with chiral fermion
generations from the heterotic string by turning on a Wilson line on a
non-simply connected Calabi-Yau threefold with an $SU(5)$ gauge group. 
For this we construct
stable ${\bf Z_2}$-invariant $SU(4)\times U(1)$ bundles
on an elliptically fibered cover Calabi-Yau threefold of special 
fibration type (the $B$-fibration). The construction makes use of
a modified spectral cover approach giving just invariant bundles.
\end{abstract}

\newpage
\section{Introduction}
Attempts to get a (supersymmetric) phenomenological spectrum 
with gauge group $G_{SM}$
and chiral matter content of the Standard model from the 
$E_8\times E_8$ heterotic string on a Calabi-Yau space $X$
started with embedding the spin connection in the gauge connection 
giving an unbroken $E_6$ (times a hidden $E_8$
coupling only gravitationally). More generally [\ref{W}], 
one can instead of the tangent bundle embed an $G=SU(n)$ bundle 
for $n=4$ or $5$,
leading to unbroken $H=SO(10)$ or $SU(5)$ of even greater 
phenomenological interest.

If there is a freely acting group $\G$ on the usually simply-connected $X$, 
one can work on $X'=X/\G$ with $\pi_1(X')=\G$ allowing a further breaking
of $H$ by turning on Wilson lines; the enhanced 
structure group $G\x \G$ leads to a reduced commutator.

\noindent
On $X'$ one turns on a ${\bf Z_2}$ Wilson line
of generator ${\bf  1_3} \oplus {\bf - 1_2}$
breaking $H=SU(5)$ to $G_{SM}$
\beqa
SU(5) \lra G_{SM}=SU(3)_c \, \x \, SU(2)_{ew} \, \x \, U(1)_Y \nonumber
\eeqa
(up to a ${\bf Z_6}$).
The Wilson line $W$ can be considered as a flat bundle on $X'$ induced 
from the ${\bf Z}_2$-cover $\rho: X\to X'$ via the given embedding 
of ${\bf Z_2}$ in $H=SU(5)$.
This gives, from\footnote{the properly normalized hypercharge $Y$ 
arises by a normalization factor $1/3$} ${\bf \bar{5}}=\bar{d}\oplus L$ 
and ${\bf 10}=Q\oplus \bar{u}\oplus \bar{e}$,
the fermionic matter content of the Standard model \beqa
SM\; fermions&=&Q\oplus L \oplus \bar{u}\oplus\bar{d} \oplus \bar{e}\nonumber\\
&=&({\bf 3},{\bf 2})_{1/3}\oplus ({\bf 1},{\bf 2})_{-1}\oplus 
({\bf \bar{3}},{\bf 1})_{-4/3} \oplus ({\bf \bar{3}},{\bf 1})_{2/3}  
\oplus ({\bf 1},{\bf 1})_2
\eeqa
From the decompositions of $ad_{E_8}$ under 
$G\x H= SU(4)\x SO(10)$ resp. $SU(5)\x SU(5)$
\beqa
{\bf 248} &=&
({\bf 4},{\bf 16})\oplus (\ov{\bf 4}, \ov{\bf 16})
\oplus ({\bf 6},{\bf 10})\oplus ({\bf 15},{\bf 1}) \oplus ({\bf 1},{\bf 45})\\
&=&({\bf 5},{\bf 10})\oplus (\ov{\bf 5}, \ov{\bf 10})
\oplus ({\bf 10},\ov{\bf 5})\oplus (\ov{\bf 10},{\bf 5})
\oplus ({\bf 24},{\bf 1}) \oplus ({\bf 1},{\bf 24})
\eeqa
one finds that for unbroken $SO(10)$ a consideration of the fundamental 
$V={\bf 4}$
is enough as all fermions, including a right-handed neutrino singlet,
sit in the ${\bf 16}\ra {\bf \bar{5}}\oplus {\bf 10}\oplus {\bf 1}$.
For unbroken $SU(5)$ also $\Lambda^2 V={\bf 10}$ has to be considered; 
but as the ${\bf 10}$ and the ${\bf \bar{5}}$ come in the same 
number of families (as also demanded by anomaly considerations)
it is enough to adjust $\chi(X,V)$ (the $V$-related matter)
to get all the Standard model fermions.

To describe explicitely the bundle we choose $X$ elliptically fibered
$\pi: X\ra B$ in a specific way. For the Hirzebruch surfaces $B={\bf F_m}$, 
$m=0,1,2$, $X$ turns out to be smooth. As $\pi_1(X)\neq 0$ one is actually 
working with a $G=SU(5)$ bundle
leading to an $H=SU(5)$ gauge group on a space admitting a free involution
$\tau_X$ (leaving the holomorphic three-form $\Omega$ invariant)
to get a smooth Calabi-Yau $X'=X/{\bf Z_2}$ over a base $B'$.

To compute the generation number one
has to work on $X$ as $X'$ does not have a section but only a bi-section 
(left over from the two sections of $X$) and so one can not use the 
spectral cover method there directly.
On $X'$ the generation number is reduced by $|\G|$.

If the bundle $V'$ over $X'$ is a 3-generation bundle
then the bundle $V=\rho^*V'$ on $X$ has $6$ generations and is 'moddable' 
by construction. Conversely, having constructed a bundle above on $X$ with 
$6$ generations, one assures that it can be modded out by $\tau_X$ 
(to get the searched for bundle on $X'$) by demanding that $V$ should be 
$\tau_X$-invariant. So one has to specify a $ \tau_X $-invariant $SU(5)$ 
bundle on $X$ that, besides fulfilling
some further requirements of the spectral cover construction (cf. below), 
leads to $6$ generations. For important work related to this question
cf. [\ref{DOPW1}], [\ref{DOPW2}], [\ref{DOPW3}], [\ref{DHOR}], [\ref{BD}] 
and literature cited there.

As it will be our goal to 'mod' not just the Calabi-Yau spaces but also the 
geometric data describing the bundle (and this transformation
of bundle data into geometric data uses in an essential way the elliptic 
fibration structure) we will search only for actions which
preserve the fibration structure, i.e., $\tau_B \cdot \pi=\pi\cdot \tau_X$ 
with $\tau_B$ an action on the base
\beqa
X & \stackrel {\tau_X}{\lra}  & X\nonumber\\
\label{involutions covering}
\pi \da & & \; \da \, \pi\\
B & \stackrel {\tau_B}{\lra} & B\nonumber
\eeqa
Therefore our elliptically fibered Calabi-Yau spaces will actually have $2$ 
sections\footnote{We will use the same notation for a section, its image and 
its cohomology class.} $\sigma_1$ and $ \sigma_2=\tau_X \si_1$
($B$-model spaces). Turning this around, if one wants to construct $ \tau_X$ 
by choosing a specific $X$, we will look for a
Calabi-Yau $X$ with a type of elliptic fibration which has besides the 
usually assumed single section ($A$-model) a second one ($B$-model);
this will then lead to a free involution $\tau_X$ on $Z$.

When investigating the invariance of $V$ we are led to consider
a version of the spectral cover method especially adapted to
the situation with two sections. Concretely we will
work with a modified spectral surface and Poincar\'e bundle 
(with $\Si=\si_1+\si_2$)
\beqa
C&=&\frac{n}{2}\Si+\eta\\
P&=&\cO\big(2\De - \Si_I-\Si_{II}-c_1\big)
\eeqa
Obviously this assumes that $V$ has {\em even} rank $n$. 
Therefore the original strategy to obtain an $SU(5)$ gauge group 
(in the observable sector) from an $SU(5)$ bundle has to be modified.
Concretely one choses an $SU(4)$ bundle (of commutator $SO(10)$ in $E_8$)
and twists this with a further invariant line bundle $\ccL$ 
(of $c_1(\ccL)=D=x\Si+\al$)
\beqa
\cV=V\ox \ccL(D)
\eeqa
The structure group is $U(4)=SU(4)\cdot U(1)_X$
but as the difference to $SU(4)\x U(1)_X$ is
only a discrete group ${\bf Z_4}$, and group-theoretical statements
are here meant on the level of Lie-algebras, we will refer to
$SU(4)\x U(1)_X$ as structure group giving $SU(5)\times U(1)_X$ as gauge group.
The anomalous $U(1)_X$ becomes massive by the Green-Schwarz mechanism.

This bundle $\cV$ is then embedded in $E_8$ as a sum of stable bundles
$\tilde{\cV}=\cV\oplus \det^{-1}(\cV)=V\ox \ccL \oplus \ccL^{-4}$
with $c_1(\tilde{\cV})=0$.
From the condition of effectivity of the five-brane $W$ in
\beqa
c_2(V)-10D^2+W=c_2(X)
\eeqa
one realises that\footnote{apart from a case, not treated in this paper, 
which has special features} one is forced to allow
for an additional twist by an invariant line bundle with $x\neq 0$.
This however leads to a problem in the Donaldson-Uhlenbeck-Yau (DUY) condition
$c_1(V)J^2=0$ where $J=\ep J_0+H_B$ is a Kahler potential for which
stability of $V$ can be guaranteed (here $H_B$ is a Kahler class on the base). 
As the concrete bound $\ep \leq \ep_*$
from which on $J$ is appropriate is not known explicitely one has
to solve the DUY equation in every oder in $\ep$ individually which leads for 
the constant term to $2xH_B^2=0$.

Therefore one must go beyond tree-level here and invoke the one-loop
correction to the DUY equation [\ref{blume}] which in turn leads to further 
conditions assuring positivity of the dilaton $\phi$ and of the
gauge kinetic function. These two inequalities taken together with the two 
inequalities assuring effectivity of $W$ and the further inequality 
assuring irreducibility (resp. ampleness) of $C$
turn out to be quite restrictive.

So we will construct rank-four vector bundles $V$
on $X$ which fulfill the following conditions \begin{itemize}
\item $\tilde{\cV}$ is $\tau_{X}$-invariant and
satisfies one-loop modified DUY condition 
\item $c_1(\tilde{\cV)} \equiv 0 \, ({\rm mod}\ 2)$, \hspace{.3cm}
$c_2(X) - \big(c_2(V)-10c_1(\ccL)^2\big)$ is effective , \hspace{.3cm}
$\chi(X,\tilde{\cV)} =N_{gen}$ \end{itemize}
Having obtained the chiral fermions of the Standard Model
one would like to count also the number of Higgs multiplets and moduli. 
This will be considered elsewhere.

In {\em section $2$} the $B$-model spaces along with their cohomological 
data are introduced. In {\em section $3$} the modified spectral cover 
construction
of bundles is introduced
and the Chern classes of $V$ and its twist are computed.
In {\em section $4$}
we are writing down for ${\bf F_0}$ and ${\bf F_2}$ a free involution 
and the $\tau_X$-action on $V$ is described in detail.
In  {\em section $5$} we give some group-theoretical details
and describe the embedding in $E_8$;
then we make the condition for the effectivity of the ensuing fivebrane 
explicit. In  {\em section $6$} the stability condition is made explicit
and the relation with the one-loop modified DUY equation is described.
Finally, in {\em section $7$} we list all numerical conditions which result 
from the analysis of these models.

B. A. is supported by DFG-SFB 647/A1. 


\section{Calabi-Yau threefold with two sections}

\resetcounter

In this paper we will consider a Calabi-Yau threefold $X$ which is 
elliptically fibered over a Hirzebruch surface ${\bf F}_m$ and whose 
generic fiber is described by the so-called $B$-fiber ${\bf P}_ {1,2,1}(4)$
instead of the usual $A$-fiber ${\bf P}_{2,3,1}(6)$ (the subscripts 
indicate the weights of $x,y,z$). $X$ is described by a generalized 
Weierstrass equation which embeds $X$ in a weighted projective space 
bundle over ${\bf F_m}$
\beqa y^2+x^4+a_2x^2z^2+b_3xz^3+c_4z^4=0
\label{weier}
\eeqa
where $x,y,z$ and $a,b,c$ are sections of $K_B^{-i}$ with $i=1,2,0$ 
and $i=2,3,4$, respectively.

$X$ admits two cohomologically inequivalent section $\si_1,\si_2$. 
For this consider eqn. (\ref{weier}) at the locus $z=0$, i.e.,
$y^2=x^4 $ (after $y\ra iy$).
One finds 8 solutions which constitute the two equivalence classes 
$ (x,y,z)=(1,\pm 1,0)$ in ${\bf P}_{1,2,1}$. We choose $y=+1$, 
corresponding to the section $\si_1$, as zero in the group law, 
while the other one can be brought, for special points in the moduli space, 
to a half-division point (in the group law) leading to the shift-involution, 
cf. section \ref{freeinvo}. Let us keep on record the relation of divisors
\beqa
(z)=\Si:=\si_1+\si_2\hspace{4cm} \si_1\cdot \si_2=0
\eeqa

One finds for the Chern-classes of $X$ (cf. [\ref{ACK}];
we use the notation $c_i=\pi^*c_i(B)$)
\beqa
\label{c2B}
c_2(X)=  6\; \Si \; c_1 +c_2+ 5c_1^2, \;\;\;\;\;\;\;\;\; \; \;  
\;\;\;\;\;\;\;\;\;\;\; c_3(X)=-36c_1^2
\eeqa
{}From the weights $a_2$, $b_3$ and $c_4$ of the defining equation 
one gets $5^2+7^2+9^2-3-3-1=148$ complexe structure deformations 
over ${\bf F_0}$. This is consistent with the Euler number and the 
$h^{1,1} (X)=4$ K\"ahler classes
\beqa
h^{1,1}(X)=4, \ \ \  h^{2,1}(X)=148, \ \ {\rm and}\ \  e(X)=-288
\eeqa
For later use let us also note the adjunction relations 
(with $\si_i:=\si_i(B)$, $i=1,2$) \beqa
\label{adj rel}
\si_i^2=-c_1 \si_i\, , \;\;\;\;\; \Si^2=-c_1\Si
\eeqa
In this paper the base $B$ of $X$
is given by a Hirzebruch surface ${\bf F_m}$ (with $m=0,2$) 
with $H^2(B, {\bf Z})$ generated by the effective base and 
fiber classes $b$ and $f$
(with the intersection relations $b^2=-m$, $b\cdot f=1$ and $f^2=0$).
$\eta=xb+yf$ effective, denoted as $\eta \geq 0$, means then 
$x\geq 0$ and $y\geq 0$.
The K\"ahler cone is described in the appendix.

\section{Bundles from spectral covers}

\resetcounter

For the description of vector bundles $V$ on $B$-fibered 
Calabi-Yau threefolds $X$ we will apply the spectral cover construction
(equivalently a relative \FM). The spectral data are an
effective divisor $C$ of $X$ ("the spectral surface") and
a line bundle $L$ on $C$. The correspondence between $(C,L)$ and $V$
was described for bundles on $A$-fibered Calabi-Yau in [\ref{FMW}], 
[\ref{FMW99}], [\ref{CD}], [\ref{an}], [\ref{C c3}]
and for the $B$-fibered case in [\ref{ACK}], [\ref{C}].

For our applications let us briefly recall some facts about the
spectral cover construction for the $A$-models. One first forms the fiber
product $X\times_B {\hat X}$ and denote the projections on $X$ and $
{\hat X}$ by $p$ and ${\hat p}$, respectively. Points $q$ on the
fiber ${\hat E_b}$ of ${\hat X}$ will parametrize degree zero line
bundles ${\cal L}_q$ on $E_b$
for each $b\in B$; as it is usual we will identify $X\iso\X$.
There exists the so-called universal Poincar\'e line bundle $\Pc$ on 
$X\fp\X=X_I\fp X_{II}$ whose restrictions to the $E_b \times q_b$
are just the ${\cal L}_q$. $\Pc$ is defined only up to a tensor
product by the pullback of a line bundle on $X_{II}$ and one can normalize
it by letting $\rest\Pc{\sigma_1\fp \X}\simeq \Oc_{X}$ leading to
$c_1(\Pc)=\De-\si_{1,I}-\si_{1,II}-c_1$.
One then considers the spectral cover surface $C\subset X$ of class 
$C=n\si_1+\eta$ which is an $n$-fold cover of $B$ and forms the fiber 
product $X\times_B C$. The
relative Fourier-Mukai transform constructs a vector bundle $V$ from
its spectral data $(C,L)$ (where ${\hat p}_C\colon X\times_B C\to C$)
\beqa
V=p_{\ast}\Big( ({\hat p}_C^{\ast}L)\otimes {\cal P} \Big)
\eeqa
We introduce now a new procedure where we make the whole bundle 
construction symmetric ($\tau$-invariant) from the beginning. 
For this we define for $even$ $n$ (with suitable pull-backs via $p$ 
and $p_{II}=\hat{p}$ understood
in (\ref{modified Poincare}); $\Delta$ is the diagonal in $X\x_B X$)
\beqa
\label{modified surface}
C&=&\frac{n}{2}\Si+\eta\\
\label{modified Poincare}
c_1(P)&=&2\De - \Si_I-\Si_{II}-c_1\;\;\;\;\;\;\;\;\;\;\;\;
\;\;\;\;\;\;\;\;\;\;\;
( \, p_*\, c_1(P)=0 \, )
\eeqa
As we will usually assume that $C$ is effective and irreducible, 
let us review the conditions which we have to impose on $\eta$ to assure this.
$C$ is effective exactly if $\eta$ is, what we denote by $\eta \geq 0$, 
and $C$ is irreducible when (cf. [\ref{BDO}])
\beqa \label{C irreducible}
\eta\cdot b\geq 0, \ \ \ \eta-\frac{n}{2}c_1\geq 0.
\eeqa

Actually we will even assume that $C$ is ample
(cf. appendix) so that the classification of line bundles
on $C$ is discrete and $L$ is determined by $c_1(L)$ up to isomorphism.
For then $h^{1,0}(C)=0$ as $h^1(C, {\cal O}_C)=h^2(X, {\cal O}_X(-C))$
by the long exact sequence associated to
$0\to {\cal O}_X(-C)\to {\cal O}_X\to{\cal O}_C\to 0$; then
$h^2(X,{\cal O}_X(-C))=h^1(X, {\cal O}_X (C))=0$ by the Kodaira vanishing 
theorem.

Applying the Grothendieck-Riemann-Roch theorem to the covering $\pi_C
\colon C\to  B$ gives
\beqa
\ch(\pi_{C*}L){\rm Td}(B)=\pi_{C* }(\ch(L){\rm Td}(C))
\label{Grr}
\eeqa
From $c_1(V)=0$ one derives thereby
the general\footnote{The space of line bundles $Pic(C)$ on $C$ is 
generically not simply characterized
by pullbacks of bundles from $X$; new divisor classes on $C$ can
appear.} expression for $L$ (with $\rho:=\mu + \frac{n}{2}\la$)
\beqa
\label{c_1(L) form}
c_1(L)&=&\frac{C+c_1}{2}+\ga\;\;\;\;\;\;\;\;\;\;\;\;
\;\;\;\;\;\;\;\;\;\;\;\;\;\;\;\;\;\;\;\;\;\;\;\;
( \, (\pi_C)_*\, \ga=0 \, )\\
\label{gamma form}
\ga&=&\la\big(n\si_1-(\eta - \frac{n}{2}c_1)\big) + \mu (\si_1-\si_2)
=\la\big(\frac{n}{2}\Si-(\eta - \frac{n}{2}c_1)\big) + \rho (\si_1-\si_2)\;\;\;
\eeqa

So for $n = 4$, and $\rho=0$ as we later have to assure invariance of $V$ (we
had chosen $c_1$ even), $c_1(L)$ is an integral class if $\la \in {\bf
Z}+\frac{1}{2}$ or if $\la \in {\bf Z}$ and $\eta$ is even.

The Grothendieck-Riemann-Roch
theorem for the covering $p:X\x_B C\ra X$ gives
\beqa
\ch(V)\; {\rm Td}(X)
&=&p_{\ast}\Big( ({\hat p}^{\ast}_C \ch(L))\ch(P) \;
{\rm Td}(X\times_BC)\Big)\\
c_2(V)&=&2\big( \eta + \frac{n}{4}c_1 \big) \Si 
-2\rho\big(\eta-\frac{n}{2}c_1\big)(\si_1-\si_2)
-\rho^2 c_1\big(\eta-\frac{n}{2}c_1\big)\nonumber\\
&&+ 2\eta c_1
+\frac{1}{2}(\la^2-\frac{1}{4})n\eta\big(\eta-\frac{n}{2}c_1\big)
-\frac{n(\frac{n^2}{4}-1)}{24}c_1^2\;\;\;\;\;\;\;\;\;\;\\
\frac{1}{2}c_3(V)&=& 4 \la \eta \big(\eta - \frac{n}{2}c_1\big) \eeqa

{\it Twisting by $\ccL(D)$}

For group-theoretical reasons explained later and for reasons of 
effectivity of the ensuing five-brane class we still have to twist
our $SU(n)$ vector bundle by an invariant line bundle $\ccL(D)$ where
$D=x\Si+\al$ to get $\cV=V\ox \ccL$
and actually then to consider $\tilde{\cV}=V\ox\ccL\oplus \ccL^{-4}$.
Now let us compute the generation number. For the double covering 
$\rho:X\ra X'$ with vector bundles $\tilde{\cV}$ and $\tilde{\cV}'$ one has
$\chi(X,\tilde{\cV})=2 \chi(X',\tilde{\cV}')$ (here $\tilde{\cV}'$ 
with $\tilde{\cV}=\rho^* \tilde{\cV}'$ exists precisely because 
$\tilde{\cV}$ is invariant).
$N_{gen}$ can be computed
from the index $\chi(X,\tilde{\cV})$. The net-number of chiral matter 
multiplets (the 'number of generations')
is given by (note that $c_1(\tilde{\cV})=0$ and $c_2=4, c_1^2=8, n=4$)
\beqa
N_{gen}&=&\chi\big(X,V\ox \ccL\oplus \ccL^{-4}\big)
=\int_X ch_3\big(V\ox \ccL\big)+ch_3(\ccL^{-4})\nonumber\\
&=&\frac{1}{2}c_3(V)-c_2(V)D-10D^3\nonumber\\
&=&4\Bigg[\la-x(\la^2-\frac{1}{4})\Bigg]\eta(\eta-2c_1)
+40x(1-4x^2)-4\al(\eta+c_1)-60x\al(\al-xc_1)\nonumber
\eeqa

\section{Involution and invariant bundles}

\resetcounter

\subsection{\label{freeinvo}Existence of a free ${\bf Z}_2$ operation}

We give a free involution $\tau_X$ on $X$ which leaves the
holomorphic three-form invariant; then $X'=X/{\tau_X}$ is a smooth Calabi-Yau.
We assume $\tau_X$ compatible with the fibration, i.e., we assume the 
existence of an involution $\tau_B$ on the base $B$ with 
$\tau_B\cdot \pi=\pi\cdot \tau_X$.

We will choose for $\tau_B$ the following operation in local (affine)
coordinates
\beqa
b=(z_1, z_2)\;\;\;
\stackrel{\tau_B}{\longrightarrow} \;\;\;
-b=\tau_B(b)=(-z_1, -z_2)
\eeqa

The idea for the construction of $\tau_X$ is to combine $\tau_B$
with an operation on the fibers. A free involution on a smooth
elliptic curve is given by translation
by a half-division point. Such an object has to exist
globally; this is the reason we have chosen the $B$-fibration
where $X$ possesses a second section. If we would tune $\si_2(b)\in E_b$
to be a half-division point the condition $b_3=0$ would ensue and $X$
would become singular. Therefore this idea has to be enhanced.
Furthermore, even for a $B$-fibered $X$ those fibers
lying over the discriminant locus in the base will be singular where
the freeness of the shift might be lost. As the fixed point locus of
$ \tau_B$
is a finite set of (four) points we can assume that it is disjoint
from the discriminant locus (so points in the potentially dangerous
singular fibers are still not fixed points of $\tau_X$; 
for discussion cf. [\ref{ACK}], [\ref{C}]).

One finds [\ref{ACK}] as $\tau_X$ over ${\bf F_m} $ with $m$ even 
(i.e., $m$ being $0$ or $2$)
the free involution
\beqa \label{iota} (z_1, z_2 \, ; x,y,z)\stackrel{\tau_X}
{\longrightarrow} (-z_1, -z_2 \, ; -x,-y,z)
\eeqa
This exchanges the points $\si_1(b)=(b \, ; 1,1,0)
$ and $\si_2(-b)=(-b\, ;  1,-1,0)$
between the fibers $E_b$ and $E_{-b}=E_{\tau_B(b)}$;
in ${\bf  P}_{1,2,1}$
the sign in the $x$-coordinate can be scaled away here in contrast to the
sign in the $y$-coordinate.
As indicated above an involution like in eqn. (\ref{iota}) could not exist on
the fiber alone, i.e.
as a map $(x,y,z)\lra (-x, -y, z)$, because this would from eqn. (\ref{weier})
force one to the locus $b_3=0$ where $X$ becomes singular
(so only then is this defined on the fiber and so, being a free
involution, a shift by a half-division point).
But it can exist combined with the base involution $\tau_B$ on a
subspace of the moduli space where the generic member is still
smooth: from eqn. (\ref{weier}) the coefficient functions should transform
under $\tau_X$ as $a_2^+, b_3^-, c_4^+$, i.e., over $ {\bf F_0}$, say,
only monomials $z_1^pz_2^q$ within $b_{6,6}$ with $p +q$ even are
forbidden;  similarly in $a_{4,4}$ and $c_{8,8}$ $p+q$ odd is
forbidden.  So the number of deformations drops to $h^{2,1}(X) = (5^2
+1)/2+(7^2-1)/2+(9^2+1)/2-1-1-1=75$. The discriminant remains generic
since enough terms in $a,b,c$ survive, so $Z$ is still smooth (cf.
[\ref{ACK}]). The Hodge numbers $(4,148)$ and $(3, 75)$ of $X$ and $X'$
show that indeed $e(X')=e(X)/2$
($X'$ has lost one divisor as the two sections are identified).

\subsection{Invariance of the bundles}

We describe conditions on the spectral data $(C,L)$ for the 
$\tau_X$-invariance of $V$.
The surface $C$ lies actually in the dual Calabi-Yau $\hat{X}$ 
where $\hat{\tau}$
operates (cf. the next subsection):
$\tilde{V}=\tau^*V$ turns out to be again a spectral cover bundle
with $\tilde{C}:=\hat{\tau}(C)$ as spectral surface.
As fiberwise semistable bundle it is fixed
up to the datum $\tilde{L}$. More precisely the argument for 
invariance now goes as follows.
The symmetric form (in $\si_1$ and $\si_2$) of $P$
suggests that  $V$ should be invariant
if $L$ is chosen also symmetric, i.e. $\rho=0$. However one has 
to take into account
that on the part of $P$ in $X_{II}=\hat{X}$, and similarly
on $L$ which sits in $X_{II}$, actually $\hat{\tau}$ operates.
The conclusion remains nevertheless correct. $\tilde{V}=\tau^*V$ 
will always be chosen to have again $C$ as its spectral surface 
(cf. next subsection) and is of the same general form as $V$, i.e., 
a spectral cover bundle of $c_1(\tilde{V})=0$,
just with different input parameters $\tilde{\la}, \tilde{\rho}$. But
these can be read off from $c_2(\tau^* V)=\tau^* c_2(V)$ 
(where the latter is the usual operation in $X$, resp. its cohomology, 
interchanging $\si_1$ and $\si_2$). This gives $\rho=0$ as necessary 
and sufficient condition for invariance.

{\em Remark on Fourier-Mukai transformation}

Usually the spectral cover construction of $V$ from $(C,L)$ is
interpreted as an equivalence of data in the framework of Fourier-Mukai 
transformations. Because of the difference between $C$ and $C_{eff}$ 
(cf.~next subsection) occuring
in our construction we will not employ the idea of an inverse transform here.
Let us nevertheless point to some facts related to the discussion above
and related invariance arguments.

$\tilde{V}=\tau^*V$ is the
Fourier-Mukai transform $FM^0$ of a {\em line bundle} 
$\tilde {L} =j^* \tilde{l}$
supported on $\tilde{C}=\hat{\tau}(C)$, where $j:\tilde{C}
\hookrightarrow X$, i.e.,
$\tau^* FM^0 i_*i^* l = FM^0 j_* j^* \tilde{l}$
where $L=i^* l$ was the line bundle datum on $C$ for $V$
(here $i:C\hookrightarrow X$).
For this recall that for $\tilde{V}$ a semistable vector bundle of
rank $n$ and degree zero on the fibers its inverse Fourier-Mukai transform
$FM^1(\tilde{V})$ is a torsion sheaf of pure dimension two on $X$ and of rank
one over its support which is a surface $j: \tilde{C}\hookrightarrow X$, 
finite of degree $n$ over $B$.
For $\tilde{C}$ smooth $FM^1(\tilde{V})=j_*\tilde{L}$ is just the extension 
by zero of some torsion free rank one sheaf which is actually a
{\em line bundle} $\tilde{L}\in Pic(\tilde{C})$:
for $\pi_{\tilde{C}*}\tilde{L}= \tilde{V}|_B$ is locally free and
$\pi_{\tilde{C}}: \tilde{C} \to B$ is a finite flat surjective morphism 
so $\tilde{L}$ is
locally free as well.

\noindent
Now for a Poincar\'e bundle which would be strictly invariant in the sense 
that $(\tau\times{\hat\tau})^*P=P$ one can indeed show that
$\tau^*FM^0(E)=FM^0(\tau^*E)$ (here $E=i_* L$). For this note that
\beqa
\tau^*FM^0(E)&=&\tau^*\pi_*(\hat{\pi}^*E\otimes P)
=\pi_*(\hat{\pi}^*(\hat{\tau}^*E)\otimes P)\nonumber\\
&=&FM^0(\hat{\tau}^*E)
\eeqa
Thus one gets invariance if $FM^0(\hat{\tau}^*E)=FM^0(E)$ for which it is
sufficient that $\hat{\tau}^*E=E$.


\subsection{Operation on the dual Calabi-Yau $\hat{X}$}

Let us see how $\tau_X$ operates effectively in the dual Calabi-Yau $\hat{X}$.
$E_b$ has the equation
\beqa
y^2=x^4+a_2^+(b)x^2z^2+b_3^-(b)xz^3+c_4^+(b)z^4
\eeqa
Similarly $E_{-b}$ with $-b:=\tau_B(b)$ has the equation
\beqa
y^2=x^4+a_2^+(b)x^2z^2-b_3^-(b)xz^3+c_4^+(b)z^4
\eeqa
$\tau$ maps $E_{-b}$ to $E_{b}$ and one gets for the transformed bundle
\beqa
V|_{E_b}=\bigoplus_{i=1}^n \cO_{E_b}\big(q_i(b)-\si_1(b)\big)
\Longrightarrow
(\tau^*V)|_{E_b}=\bigoplus_{i=1}^n \cO_{E_b}\big(\tau  q_i(-b) - \si_2
(b)\big)
\eeqa
As $\cO_{E_b}\big(\tau q_i(-b) - \si_2(b)\big)\cong
\cO_{E_b}\big(t_{\si_2(b)}^{-1}\tau q_i(-b) - \si_1(b)\big)$ where
$t_ {\si_2(b)}$ is the translation in the group law
one finds that $\tau^* V = V$ amounts fiberwise to the relation
$\{q_i(b)\}=\{ t_{\si_2(b)}^{-1}\tau q_i(-b) \}$, i.e., to
$t_{\si_2}^{-1}\tau C=C$. So $\hat{\tau}:=t_{\si_2}^{-1}\circ \tau $
is
the relevant operation $\hat{\tau}$ on $\hat{X}$ \beqa
\tau_X = t_{\si_2}\circ\hat{\tau}
\eeqa
We did not assume
that $\si_2(b)$ is a half-division point (so $t_{\si_2}$ is not an
involution); if one would do so by tuning $b_3=0$ the space $X$
would become singular and only then acts $t_{\si_2}$ as 
$(b;x,y,z)\ra (b;-x,-y,z)$.

{\em Actions in coordinates and in the group structure}

\noindent
Involutions $\tau_X$ covering $\tau_B$ (in the sense of eqn.
(\ref{involutions covering})) are determined by involutions
$\al$ (which turns out to be $\hat{\tau}$ in our case) covering $\tau_B$ with
$\al \, \si_{1/2}=(-1) \, \si_{1/2}$
(the minus refers to the inversion in the group law). (Cf. also [\ref{DOPW3}].)

Now let us define a decomposition of $\tau$ in the coordinate involutions \beqa
\tau = \iota \circ \beta
\eeqa
with ($\iota$, covering ${\bf 1_B}$, is fiberwise the covering involution
for the obvious map to ${\bf P^1_{x,z}}$)
\beqa
\iota: \big(b; x,y,z\big) \lra  \big(b; x,-y,z \big)
\;\;\;\; , \;\;\;\;\;\;\;\;\;
\beta: \big(b; x,y,z\big) \lra  \big(-b; -x,y,z\big)
\eeqa
($\beta$ keeps $\si_i$ fixed).
One has the following relation between coordinate map
and the structural maps
(note that both sides just act on the fiber and
interchange the $\si_i$)
\beqa
\label{decomposition}
\iota = t_{\si_2}\circ (-1)
\eeqa
First note here that $\iota$ is independent of the element 
$q$ chosen as group zero;
similarly the translation becomes in general 
$(t_{\si_2})_q=t_{\si_2}\circ t_q^{-1}$ and the inversion
$(-1)_q=t_q\circ (-1)$, so the right hand side is
independent of the zero chosen.

Concerning the proof of eqn. (\ref{decomposition}) note that a
holomorphic map on the fiber ${\bf C}/\Lambda$ is $t_q\circ \rho$ with
$\rho$ a group homomorphism, $t_q$ a translation (here $q$ must
be $\si_2$). $\rho$ lifts to a linear transformation
$z\ra az+b$ of ${\bf C}$ ($b\in \Lambda$), so $a$, keeping
invariant the lattice $\Lambda$, is in ${\bf Z}$
(${\bf C}/\Lambda$ in general has no complex multiplication); $\rho$ is an
isomorphism as $\iota$ is, so $a=\pm1$ and $a=-1$ as $\rho \si_2 =
(-1) \si_2$.
\beqa
\tau = \iota \circ \beta = t_{\si_2}\circ \underline{(-1)\circ \beta }
=t_{\si_2}\circ \underline{\hat{\tau}}
\eeqa

{\em Application to the spectral points}

\noindent
For the choice of our Poincare bundle in
$V=p_*\big( p_C^*L \ox P\big)$ one finds
\beqa
P|_{E\x q}&=&\ccL\Big( (2q-\si_2)-\si_1\Big)
\eeqa
giving effective spectral points $C^{eff}=\{2q-\si_2|q\in C\}$ 
($"2"$ and $"-"$ refer to the group).

What we actually want to achieve is that
$\hat{\tau}C^{eff}=C^{eff}$ as the $\hat{\tau}$-invariance condition 
concerns the points $q_i^{eff}$ corresponding to line bundle summands 
on the fiber.
$\beta$, being holomorphic and sending $\si_1$ to itself, is
a group homomorphism between respective fibers. With $2q-\si_2=q-\iota q$ 
(operations in the group) from eqn. (\ref{decomposition}), 
one gets as condition
\beqa
(-1)\beta (1-\iota)q_i(b)=(1-\iota)q_i(-b)
\eeqa
As $(-1)$ and $\beta$ are group homomorphisms this will be fulfilled
if we can assure that \beqa
\label{effective invariance}
q_i(-b)=\iota \beta q_i(b)
\eeqa
That is, the reduction achieved amounts to (note that fiberwise
$C^{eff}=(1-\iota)C$)
\beqa
\iota \; \beta \; C=C\;\;\;\;
\Lra \;\;\;\; (-1)\; \beta \; C^{eff}=C^{eff}
\eeqa

{\em The equation of the spectral cover surface}

To show how this condition can be implemented we give the coordinate 
description of our spectral cover surface $C$. In the appendix 
we recall the corresponding relations in the $A$-model
elliptic fibration and derive the expression for our case.

{}From there one gets the equation of a spectral cover surface $C$ (for $n=4$)
\beqa
w = \al_{20}x^2+\al_{02}y+\al_{10}xz+\al_{00}z^2 = 0 \eeqa
This shows that invariance of $C$ under $\beta$ would mean
that the coefficient functions $\al_{ij}$ transform invariantly under
$\tau_B$ except $\al_{10}$ which should transform
anti-invariantly (or the other way around); similarly invariance of
$C$
under $\iota \beta$ as in eqn. (\ref{effective invariance}) (to assure the
necessary $\hat{\tau}=(-1)\beta$-invariance of $C^{eff}$) means
that $\al_{02}$ and $\al_{10}$ have to transform with the other sign
then $\al_{20}$ and $\al_{00}$, a condition which can easily be imposed.


\section{The $E_8$ embedding and the massive $U(1)$}

\resetcounter

We will explain now more precisely our strategy outlined in the 
introduction how to get an $SU(5)$ GUT group and
describe the embedding of the structure  group into $E_8$.

The spectral cover construction leads in general to $U(n)$ bundles $\cV$
(the 'non-split case') with $c_1(\cV)=\al$, cf. [\ref{AH}].
For us it will enough to consider the 'split case'
\beqa
\cV=V\otimes \ccL(D)
\eeqa
where $V$ is an $SU(n)$ bundle and $\ccL(D)$ is a line bundle
of $c_1(\ccL(D))=D=x\Si+\al$. So $c_1(\cV)\equiv 0 \, (n)$ 
as $D$ is integral. Conversely,
if $c_1(\cV)\equiv 0 \, (n)$,
one can split off an {\em integral} class $D$ of $c_1(\cV)=nD$
and define a corresponding line bundle $\ccL(D)$ such that 
$V:=\cV\otimes \ccL(-D)$ is an $SU(n)$ bundle,
i.e., one can think of $\cV$ then as $V\ox\ccL(D)$.

Note that the structure group $U(n)$ arises in this case from 
$SU(n)\cdot U(1)$ (the latter factor is understood here always 
as embedded by multiples of the identity matrix)
whereas for a bundle $V\oplus \ccL(D)$ the structure group would 
be the direct product $SU(n)\times U(1)$. Note that there is a 
morphism  $f\colon SU(n)\times U(1) \to U(n)$
sending  $(a,b)\mapsto a\cdot b$. The image of this morphism is $U(n)
=SU(n) \cdot U(1)$, so $SU(n)\cdot U(1) = \Big( SU(n) \times U(1)
\Big)/{\rm ker}(f)$. The subgroup ${\rm ker}(f)$ is formed by all
pairs $(\lambda \cdot {\rm Id_n}, \lambda^{-1})$ where $\lambda \in
{\bf C}$ with $\lambda \cdot {\rm Id_n} \in SU(n)$, i.e.,
$\lambda^n=1$ and ${\rm ker}(f) = {\bf Z_n}$ (the group of 
$n$-th roots of unity). As the difference between the direct
product and the product is just a discrete group,
and since all group theoretical statements in this
paper are understood on the level of Lie-algebras, we will write 
$SU(4)\times U(1)$ instead of $SU(4)\cdot U(1)$ for our structure group $G$.

Let us make the embedding of $G$ in $E_8$ more explicit.
One embeds a $U(4)$ bundle block-diagonally via
$U(4)\ni A \lra \tiny{\left( \begin{array}{cc} A&0\\ 0&\det^{-1}A 
\end{array}\right)}
\in SU(5)$.
In our case $\cV=V\ox \ccL$ with $c_1(\cV)=4c_1(\ccL)$, so in this sense
one works actually with the bundle $\tilde{\cV}=\cV\oplus \ccL^{-4}$.

The commutator of $G=SU(4)\times U(1)_X$ in $E_8$
is given by $H=SU(5)\times U(1)_X$, the observed gauge group.
The adjoint representation of $E_8$ decomposes under 
$SU(4)\times SU(5)\times U(1)_X$ as follows
(with $ad(E_8)=\bigoplus_i U_i^{SU(4)}\otimes R_i^{SO(10)}=
\bigoplus_i (U_i, R_i)=\bigoplus_i (U_i, S_i^{SU(5)})_{t_i^{U(1)}}$)
\beqa
\label{break}
{\bf 248} &\stackrel{SU(5)\times SU(5)}{\longrightarrow} &
({\bf 5},{\bf 10})\oplus (\ov{\bf 5},\ov{\bf 10})\oplus 
({\bf 10}, \ov{{\bf 5}})\oplus (\ov{\bf 10}, {\bf 5})
\oplus ({\bf 24},{\bf 1}) \oplus ({\bf 1},{\bf 24})\\
\label{multiplet decomposition}
&\stackrel{SU(4)\times SU(5)\times U(1)_X}{\longrightarrow} &
\hskip-0.3cm
\Bigg(({\bf 4},{\bf 1})_{-5} \oplus ({\bf 4}, \ov{\bf 5})_3 \oplus
({\bf 4},{\bf 10})_{-1}\Bigg)
\oplus \Bigg( (\ov{\bf 4},{\bf 1})_{5} \oplus (\ov{\bf 4}, {\bf 5})_{-3}
\oplus (\ov{\bf 4},\ov{\bf 10})_{1}\Bigg) \nonumber \\
&&\oplus ({\bf 6},{\bf 5})_2 \oplus ({\bf 6},\ov{\bf 5})_{-2}\nonumber \\
&&\oplus ({\bf 15},{\bf 1})_0 \oplus ({\bf 1}, {\bf 1})_0 
\oplus ({\bf 1},{\bf 10})_4 \oplus
({\bf 1}, \ov{\bf 10})_{-4} \oplus ({\bf 1}, {\bf 24})_0 \nonumber
\eeqa

The $SU(5)$ representations are given as an auxiliary step.
The full decomposition, identical to an auxiliary $SU(4)\x SO(10)$ step,
leads to the right-handed neutrino $\nu_R$.\\
The massless (charged) matter content is $\bigoplus (S_k)_{t_k} 
={\bf 1_{-5}} \oplus {\bf \bar{5}_3} \oplus {\bf 10_{-1}}
\oplus {\bf \bar{5}_{-2}}\oplus {\bf 10_4}$; one can write conditions
for the absence of net generations of the
exotic matter given by the last tow summands;
additionally (besides the gauge bosons 
$({\bf 1}, {\bf 1})_0 \oplus ({\bf 1}, {\bf 24})_0$ of $H$)
some neutral matter given by singlets (moduli) arises from $End(V)$, 
i.e., $({\bf 15},{\bf 1})_0$.

Precisely those $U(1)$'s in $H$ which occur already in $G$ 
(a so-called $U(1)$ of type I; other $U(1)$'s in $H$ 
are called to be of type II)
are anomalous [\ref{Distler}], [\ref{Dine}], [\ref{Lukas}], [\ref{blume}].
The anomalous $U(1)_X$ can gain a mass by absorbing some of the
would be massless axions via the Green-Schwarz mechanism, that is,
the gauge field is eliminated from the low energy spectrum by
combining with an axion and so becoming massive. One has to check
that the anomalies related to $U(1)_X$ do not cancel accidentally
(i.e., that the mixed abelian-gravitational, the mixed abelian-non-
abelian and the pure cubic abelian anomaly do not all vanish).
Computing the anomaly-coefficients of $U(1)_X$ we find (cf. [\ref{blume}])
\beqa
A_{U(1)-G_{\mu\nu}^2}&=&\sum tr_{(S_k)_{t_k}}q \, \cdot 
\chi(X, U_k \ox t_k)\nonumber\\
&=&10 D \cdot \Big( 12 \cdot \big(-c_2(V)+10D^2\big) + 5c_2(X)\Big)\\
A_{U(1)-SU(5)^2}&=&
\sum q_{t_k} C_2(S_k)\, \cdot \chi(X, U_k \ox t_k)\nonumber\\
&=&10 D \cdot \Big(2\big(-c_2(V)+10D^2\big) + c_2(X)\Big)\\
A_{U(1)^3}&=&
\sum tr_{(S_k)_{t_k}}q^3 \, \cdot \chi(X, U_k \ox t_k)\nonumber\\
&=&200 D \cdot \Big( 6  \big(-c_2(V)+10D^2\big)+40D^2+ 3c_2(X)\Big)
\eeqa
(with $C_2$ normalised to give $C_2(\bar{f})=1$, $C_2(\Lambda^2 f)=3$ 
for $SU(5)$).
For $x\neq 0$ (the case of interest for us, cf. below) the last
two conditions are not proportional as $D^3\neq 0$. In any case the 
first condition is independent, so not all three coefficients will vanish.

\subsection{Effectivity of the fivebrane}

In the presence of magnetic five-branes the Bianchi identity 
$H$-field reads
\beqa
dH=\tr R\wedge R-\frac{1}{30}\Tr F\wedge F +\sum_{\rm 5-branes}\delta_5^{(4)}
\label{bian}
\eeqa
where $R$ and $F$ are the associated curvature forms of the spin
connection on $X$ and the gauge connection on $\cV$ and the last term
is the source term contributed by the five-branes. Here $\tr$
refers to the trace of the endomorphism of the tangent bundle of $X$ 
and $\Tr$ denotes the trace in the adjoint representation of $G$.

This leads in our case of \beqa
\tilde{\cV}=\cV \oplus \ccL^{-4}
\eeqa
to the anomaly equation
\beqa
-ch_2(X)=-ch_2(\cV)-ch_2(\ccL^{-4})+W
\eeqa
This gives with $c_1(\cV)=-c_1(\ccL^{-4})$ and $c_1(\ccL)=D$
\beqa
c_2(X)&=&c_2(\cV)-c_1^2(\ccL^{-4})+W\nonumber\\
&=&c_2(V)-10D^2+W
\eeqa
Note that the last term in eqn.(\ref{bian}) is formally a
current that integrates to one in the direction transverse to a
five-brane whose class we denote by $W$. The class $W$ is
the Poincar\'e dual of the sum of all sources and represents a class in 
$H_2(X,{\bf Z})$. Supersymmetry demands $W$ to be the class of an effective 
curve in $X$. For simplicity we will think of the curve of $W$ as being 
irreducible.

We have a factorization $\cV=V\otimes \ccL(D)$
where $V$ is an $SU(n)$ bundle and $\ccL(D)$ is an line bundle with 
$c_1(\ccL(D))=D$. With the decomposition of $W$
\beqa
W=W_B+W_F=w_B \, \Si + a_f F
\eeqa (where $W_B$ is an effective curve class and $a_f\geq 0$) one finds
\beqa
W&=& c_2(X)-c_2(V)+10D^2\\
&=&\Big(6c_1\Si +c_2+5c_1^2\Big)
-2(\eta +\frac{n}{4}c_1)\Si -kF + 10D^2\nonumber\\
&=&\Bigg(6c_1-2(\eta+\frac{n}{4}c_1)+10x(2\al-xc_1)\Bigg)\Si
+\Bigg(c_2+5c_1^2-kF+10\alpha^2\Bigg)\nonumber
\eeqa
giving the effectivity conditions (with $c_1^2=8$, $c_2=4$, $n=4$) \beqa
\label{base positivity twisted rev}
w_B&=&4c_1-2\eta+10x(2\al - xc_1) \geq 0\\
a_f&=&48-2(\la^2-\frac{1}{4})\eta(\eta -  2c_1)
-2\eta c_1 +10\al^2 \geq 0
\eeqa
Note that (\ref{base positivity twisted rev}) amounts to $\eta\leq 2 c_1$
for $x=0$, i.e., to $\eta=2c_1$ where $C$ is not ample and $h^{1,0}(C)\neq 0$.

\section{The DUY constraint and its 1-Loop Modification}

\resetcounter

Like the Calabi-Yau condition on the underlying space $X$, the 
holomorphicity and stability of the vector bundle $V$ are direct 
consequences of the required four-dimensional supersymmetry. 
The demand is that a connection $A$ on $V$ has to satisfy 
(at string tree level) the Donaldson-Uhlenbeck-Yau (DUY) equation
($J$ denotes a K\"ahler form on $X$)
\beqa
F_A^{2,0}=F_A^{0,2}=0, \quad F_A^{1,1}\wedge J^{2}=0
\label{hym}
\eeqa
The first equation implies the holomorphicity of $V$; the second 
equation is the  Hermitian-Yang-Mills (HYM) equation 
$F_A^{1,1}\wedge J^{n-1}=c\cdot I_F\cdot J^n$ 
(for $n=3$ with $c\in {\bf C}$ vanishing)
with the integrability condition
(condition for the existence of a unique solution in case 
$V$ is polystable, i.e.,
a sum of $\mu$-stable bundles with the same slope)
[\ref{UhYa}], [\ref{Donald}] \beqa
\int_X c_1(V)\wedge J^2=0
\label{duy}
\eeqa
$V$ is called $\mu$-stable with respect to some
K\"ahler class  $J$ if
its slope $\mu(V)=\frac{1}{rk(V)}\int c_1(V)\wedge J^2$ is bigger than the
slope of each subbundle $V'$ of smaller rank.

If $C$ is irreducible then $V$ will be stable for sufficiently small $\epsilon$
with respect to [\ref{FMW99}] \begin{equation}
J=\epsilon J_0+\pi^\ast H_B, \qquad {\epsilon} > 0
\label{pola}
\end{equation}
Here $J_0=x_1\si_1+x_2\si_2+h$ with $x_1+x_2>0$ and $H_B$ 
is an ample divisor in $B$. So the volume
of the fiber $F$ of $X$ is kept arbitrarily small compared to volumes of
effective curves in the base.
For details on the stability properties and generalizations we
refer to [\ref{FMW99}, \ref{Bjoern}]. Note that working on $B$-fibered
Calabi-Yau threefolds does not effect the stability proof of [\ref{FMW99}]. 
Thus our bundles are stable with respect to the K\"ahler class in 
eqn. (\ref{pola}).

The bundle $\V=V\ox \ccL(D)$ is stable with respect to 
$J=\epsilon J_0+H_B$ with
$\epsilon$ small positive, $H_B\in {\cal C}_B$ and $J_0=x_1\si_1+x_2\si_2$
(with ${\cal C}_B$ the K\"ahler cone of $B$).
The DUY constraint is
\beqa
\label{DUY constraint new}
0=DJ^2
=-\epsilon^2c_1(\al - xc_1)\Big(\sum x_i^2\Big) +2\epsilon \big(\sum x_i \big)
H_B (\al -xc_1)+2xH_B^2
\eeqa
If one tries to argue that the K\"ahler moduli in $J$ are tuned such 
that $DJ^2=0$, one encounters the difficulty
that the relevant $\epsilon$ (for which stability is known) depends itself
on the chosen $J_0$ and $H_B$. So one rather has to
make vanish the coefficients of the three different $\epsilon$-powers 
individually, so that $DJ^2=0$ independently of
the actual, unknown value of $\epsilon$. But the constant term gives
$x=0$ for which eqn. (\ref{base positivity twisted rev}) can not be relaxed.

For a rank $n$ bundle $\cV=\oplus \cV_i$ composed of $U(n_i)$ bundles of
slopes $\mu_i=\frac{1}{n_i}\int J^2 c_1(\cV_i)=:\mu$ 
(they must coincide for $\cV$ to be polystable)
one finds $\int J^2 c_1(\cV_i)=0$ for all $i$ as
$0=\int J^2 c_1(\cV)=\sum r_i \mu_i=\mu n$.
For us, having $c_1(\cV_i)=\pm 4D$, this means $\int J^2 D=0$.

{\em The 1-loop modification}

We discuss now an approach which leads to a rank
5 physical bundle for a split extension and {\em non-vanishing} $x$. This
causes the slope to be non-zero and so one has to invoke the quantum
corrected version of it. This will make the slope vanish at the one loop
level and fix the dilaton.

From the condition of effectivity of the five-brane $W$ in
$c_2(V)-10D^2+W=c_2(X)$
one realises that\footnote{apart from a 'separation case', 
which has special features} 
one is forced to allow
for an additional twist by an invariant line bundle with $x\neq 0$.
This however leads to a problem in the Donaldson-Uhlenbeck-Yau (DUY)
condition
$c_1(V)J^2=0$ where $J=\ep J_0+H_B$ is a Kahler class for which
stability of $V$ can be guaranteed (here $H_B$ is a Kahler class on the
base). As the concrete bound $\ep \leq \ep_*$
for which $J$ is appropriate is not known explicitely one has
to solve the DUY equation in every oder in $\ep$ individually which leads
for the constant term to $2xH_B^2=0$.

Therefore one must go beyond tree-level here and invoke the one-loop
correction to the DUY equation [\ref{blume}] which in turn leads to further
conditions assuring positivity of the dilaton $\phi$ and of the
gauge kinetic function. These two inequalities taken together with the two
inequalities assuring effectivity of $W$ and the further inequality assuring
irreducibility (resp. ampleness) of $C$ turn out to be quite restrictive.

So we invoke the one-loop correction of the $DUY$ integrability
constraint [\ref{blume}],[\ref{Timo}], [\ref{Blumenhagen06}]
(here $z\in [-\frac{1}{2},+\frac{1}{2}]$ refers to the position of 
$W$ in the interval
between the two $E_8$-walls)\footnote{we choose $z=0$ to illustrate; 
to determine $z_W$, open membrane instanton effects have to be included}
\beqa
0&=&\int_X c_1(\ccL)J^2-\phi\int_X c_1(\ccL)\Big(-c_2(V)
+10c_1(\ccL)^2+\frac{1}{2}c_2(X)-(\frac{1}{2}- z)^2 W  \Big)\\
&=&DJ^2 -\phi \;
D \Big[ \big( 1 - (\frac{1}{2}-z)^2\big) W-\frac{1}{2}c_2(X)\Big]
\eeqa
(corresponding to a deformed $DUY$ equation)
with $\phi=l^4_sg^2_s$ giving \beqa
\label{phi fix new rev}
\phi=\frac{DJ^2}{D\Big(\frac{3}{4}W-\frac{1}{2}c_2(X)\Big)}=
\frac{\cO(\epsilon)+xH^2_B}{(\al - x c_1)(\frac{3}{4}w_B-3c_1)
+x(\frac{3}{4}a_f-22)}>0\;\;\;\;\;\;\;\;
\eeqa
as condition.
So it is enough to have (for $x>0$; otherwise one has the reverse inequality)
\beqa
(\al - x c_1)(\frac{3}{4}w_B-3c_1)+x(\frac{3}{4}a_f-22)>0
\eeqa
The positivity of the one-loop corrected gauge kinetic function
requires [\ref{blume}],[\ref{Timo}], [\ref{Blumenhagen06}]
\beqa
0<J^3-3\phi\Big[J\Big(W-\frac{1}{2}c_2(X)-(\frac{1}{2}- z)^2 W \Big)\Big]
=\cO(\epsilon)-3\phi \Big[ \big(1- (\frac{1}{2}- z)^2 \big)W 
-\frac{1}{2}c_2(X)\Big]H_B\nonumber
\eeqa
As stability will hold for all small enough $\epsilon$ it will be
enough to show \beqa
\Big( \frac{3}{4}\big[ 4c_1-2\eta +10x(2\al -xc_1)\big]
-3c_1 \Big)H_B<0
\eeqa

\section{Model constraints and Conclusions}

\resetcounter

Let us collect all conditions on the bundle.
Note that $\la$ can be integral or half-integral if $\eta$ is even
(actually $\al$ could be half-integral by considering genuine $U(n)$ bundles $\cV$).

We obtain the equation for the generation number of $\tilde{\cV}
=V\ox \ccL\oplus\ccL^{-4}$
with $D=x\Si+\al$
\beqa
N_{gen}=4\Bigg[\la-x(\la^2-\frac{1}{4})\Bigg]\eta(\eta-2c_1)
+40x(1-4x^2)-4\al(\eta+c_1)-60x\al(\al-xc_1)
\eeqa
One notes that $N_{gen}$ is manifestly divisible by 4, except perhaps the
first term. For $n=4$ and $\rho=0$ is $\la$ either ${\bf Z}+\frac{1}{2}$ or
$\la \in{\bf Z}$ with $\eta $ even. It remains to discuss the term
$\eta(\eta-2c_1)$. If $\la\in\frac{1}{2}+{\bf Z}$ this term is divisible by
two; if $\la\in{\bf Z}$ as    $c_1$ is even and $\eta\equiv c_1\ {\rm mod\
2}$ this term is divisible by $4$. In total, also the first term is
divisible by 4. 

Therefore the physical generation number $N_{gen}/2$
downstairs on $X/{\bf Z_2}$ 
turns out to be even.

Furthermore one obtains five inequalities as conditions: 
one for the irreducibility of $C$
(this is on ${\bf F_0}$; when checking effectivity and irreducibility of $C$
note that $\eta \geq 0$ implies on ${\bf F_0}$
also $\eta \cdot b \geq 0$); actually for $C$ ample one gets a slightly 
sharper inequality; then two
for the effectivity of $W$ and two from the one-loop considerations, 
concerning
positivity of $\phi$ (where (\ref{phi rev}) is meant for $x>0$; 
otherwise one has the reverse inequality)
and of the gauge kinetic term\footnote{From (\ref{w_B z=0}) and
(\ref{gauge kin pos rev z=0})
one may try to choose $0\leq w_B\leq 4c_1$ (with $w_B\neq 4c_1$).
But this is unnecessarily restrictive: (\ref{gauge kin pos rev z=0}) 
is satisfied {\em for all} ample $H_B$ for $w_B-4c_1\leq 0$ 
(with $w_B\neq 4c_1$); but if $w_B-4c_1=(s,t)$ with already 
only one of them negative
then one can find {\em a certain} $H_B$
for which (\ref{gauge kin pos rev z=0}) holds.}
\beqa
\label{irreducibility}
\eta &\geq& 2c_1 \\
\label{w_B z=0}
w_B&=&4c_1-2\eta+10x(2\al - xc_1) \geq 0\\
a_f&=&48-2(\la^2-\frac{1}{4})\eta(\eta -  2c_1)
-2\eta c_1 +10\al^2 \geq 0\\
\label{phi rev}
\frac{1}{2}D\Big[ \frac{3}{4}W-\frac{1}{2}c_2(X)\Big]
&=&(\al - x c_1)(\frac{3}{4}w_B-3c_1)+x(\frac{3}{4}a_f-22)>0\;\;\;\;\;\;\;\;\\
\label{gauge kin pos rev z=0}
\frac{1}{2}H_B\Big[ \frac{3}{4}W-\frac{1}{2}c_2(X)\Big]
&=&\frac{3}{4}\Big( -2\eta +10x(2\al -xc_1) \Big)H_B<0
\eeqa

Therefore we have constructed invariant bundles on $B$-fibered Calabi-Yau
spaces. Thereby we can get downstairs on $X/{\bf Z_2}$ a number of $N_{gen}/2$ 
net generations of chiral fermions of the Standard model (plus an 
additional right-handed neutrino).


\appendix

\section{Appendix}

\resetcounter

\subsection{Equation of the spectral cover}

Recall that in the $A$-model with elliptic curve $zy^2=4x^3-g_2x-g_3$ 
in ${\bf P^2}={\bf P_{1,1,1}}$ with zero point $p=(0,1,0)$ the
$(z)=l$ in ${\bf P^2}$ becomes $(z)|_E=3p$ on $E$. To encode $n$ points 
on $E$ one chooses a homogeneous polynomial $w^{(hom)}_{n/2} (x,y,z)$ 
of degree $n/2$. One realises that from
its $3 n/2$ zeroes on $E$ (after Bezout's theorem) only $n$, say 
$q_i$, carry information as $n/2$ of them are always at $p$: 
for the rewriting $w^{(hom)}_{n/2}(x,y,z)=z^{n/2}w^{aff}_{n/2}(x/z,y/z)$ 
shows manifestly
$3 n/2$ zeroes at $p$ from the $z$-power and $n$ poles at $p$ and 
$n$ zeroes at the $q_i$ (note that $x/z$ and $y/z$ have a pole at 
$p$ of order $-(1-3)=2$ and $-(0-3)=3$, resp., so
the meromorphic function $w^{aff}|_E$ has that divisor).
Concretely for  $n=4$
\beqa
w^{(hom)}(x,y,z)=a_4x^2+a_3yz+a_2xz+a_0z^2
=z^2\; w^{aff}(\frac{x}{z}, \frac{y}{z})
\eeqa
This gives on $E$ for the divisor of $w^{(hom)}=w$
resp. for the divisor of zeroes of $w^{aff}$
\beqa
(w|_E)=\frac{n}{2}\sigma + \sum q_i\;\;\;\; , \;\;\;\;
\;\;\;\; (w^{aff}|_E)_0=\sum q_i
\eeqa
Globally, as $(x,y,z)$ have ${\cal L}=K_B^{-1}$-weights
$(2,3,0)$, one has $a_i\in H^0(B, {\cal O}(\eta - i c_1))$.

Now in the $B$-model with the elliptic curve in ${\bf P_{1,2,1}}$
one has the group zero $p_1=(1,1,0)$ and the point $p_2=(1,-1,0)$; let
$(z)|_E=p_1+p_2=:P$. The well-defined meromorphic functions
$x/z$ and $y/z^2$ on $E$ have polar divisors $P$ and $2P$, respectively.
For $n=4$
\beqa
w_{n/2}^{aff}(\frac{x}{z}, \frac{y}{z^2})=
\al_{20}(\frac{x}{z})^2 + \al_{02} \frac{y}{z^2}+\al_{10}\frac{x}{z}+
\al_{00}
\eeqa
From the polar divisor of the meromorphic function $w^{aff}|_E$
one reads off its total divisor\footnote{The divisor 
$2(\si_1+\si_2)-\sum^4 q_i$ (in each fiber)
of the meromorphic function $w/z^2$ gives the relation $\sum^4 q_i=2
(\si_1+\si_2)$ or $\sum^4 (q_i-\si_1)=2(\si_2-\si_1)$
in the divisor class group, or $\sum^4 q_i=2\si_2$ in the group
structure; so $\sum^4 (2q_i-\si_2)=\sum^4
q_i^{eff}=0$, i.e. one has fiberwise an $SU(n)$-bundle, cf. section 4.3.} 
$(w^{aff}|_E)=2P-\sum^4 q_i$.
So the corresponding homogeneous polynomial $w^{(hom)}=w$
\beqa
\label{C_B equ}
w_{n/2}^{(hom)}(x,y,z)=\al_{20}x^2 + \al_{02} y +\al_{10}xz+\al_{00}z^2
= z^2 w_{n/2}^{aff}
\eeqa
has on $E$ just the divisor given by the four zeroes $q_i$ 
(as the double zeroes at the $p_i$ of the $z$-power cancel 
with the double poles there of $w^{aff}$), i.e.
\beqa
(w|_E)=\sum q_i = (w^{aff}|_E)_0
\eeqa
(So this case is simpler than the $A $-model as no $n/2$ 
further zeroes at $p$ are carried along.)

Concerning the ${\cal L}=K_B^{-1}$-weights of the $\al_{ij} $, 
i.e. the transformation properties of the coefficient functions 
along the base of the fibration, $x,y,z$ have ${\cal L}$-weights $1,2,0$ 
(do not confuse them with their indicated
${\bf P_{1,2,1}}$ weights), so  
$\al_{ij}\in H^0\Big(B, {\cal O} \big (\eta-(i+j)c_1\big)\Big)$.

\subsection{The K\"ahler cone}

Concerning the base surface $B={\bf F_m}$ note that the base and fiber
classes $b$ and $f$ represent actual curves. The {\em effective cone}
(non-negative linear combinations of classes of actual curves)
is given by the condition $p\geq 0, q\geq 0$ on $\rho =pb+qf$;
this we denote by $\rho \geq 0$. The {\em K\"ahler cone} ${\cal C}_B$ of $B$
(where $\rho \in {\cal C}_B$ means $\rho \zeta > 0$ 
for all actual curves of classes $\zeta$
or equivalently $\rho b>0, \rho f>0$) is given by 
${\cal C}_B=\{t_1 b^+ + t_2 f| t_i >0 \}$ (with $b^+=b+mf$).
For example on ${\bf F_2}$ one has $c_1\notin {\cal C}_B$ as $c_1b=0$.

Let $J=x_1\si_1+x_2\si_2+H$ be an element
in the K\"ahler cone ${\cal C}_X$ ($H \in {\cal C}_B$).
Then demanding that its intersections with the curves $F$ and $\si_i \alpha$
are non-negative amounts to
\beqa
\label{first linear}
x_1+x_2 & > & 0 \\
\label{second linear}
(H-x_ic_1) \alpha & > & 0
\eeqa
Similarly intersecting $J^2$ with $\si_i$ and $\alpha$ and 
building also $J^3$ gives
\beqa
(H-x_ic_1)^2 & > & 0 \\
\label{second quadratic}
\Big( 2 \sum x_i \; H - \sum x_i^2 \; c_1 \Big) \alpha & > & 0 \\
\label{J^3}
\sum x_i \, (H-x_i c_1)^2 + \Big( 2 \sum x_i \; H - \sum x_i^2 \; c_1
\Big) H & > & 0
\eeqa
{}From this one gets\footnote{(\ref{J^3}) is not independent 
as is clear for $x_1$ and $x_2$ individually non-negative, 
and follows in general with (\ref{first linear})
(as the left hand side of (\ref{J^3}) is 
$\sum x_i (3H^2 - 3x_i Hc_1 +8x_i^2) \geq 8 \sum x_i^3$ 
using (\ref{second linear}) for $\alpha=H$).
(\ref{second quadratic}) follows from (\ref{second linear}),
giving $\Big( \sum x_i \; H - \sum x_i^2 \; c_1 \Big) \alpha \geq 0$,
with (\ref{first linear}), giving $\sum x_i H \alpha \geq 0$.}
the condition for $J$ to be ample (positive)
\beqa
\label{Kahler cone condition}
J=x_1\si_1+x_2\si_2+H\in {\cal C}_X \;\;\; 
\Longleftrightarrow \;\;\; x_1+x_2>0\;\;\; , \;\;
H-x_ic_1 &\in & {\cal C}_B \eeqa

So the condition for $C=\frac{n}{2}\Si+\eta$ to be 
{\em ample} (not only effective, cf. (\ref{C irreducible}))
is
\beqa
\label{C positive}
\eta - \frac{n}{2}c_1 \in {\cal C}_B
\eeqa

\section*{References}
\begin{enumerate}

\item
\label{W}
E. Witten, {\em New issues in manifolds of $SU(3)$ holonomy}, 
Nucl. Phys. {\bf B268} (1986) 79.

\item
\label{FMW}
R. Friedman, J. Morgan and E. Witten, {\em Vector bundles and $F$-theory},
Comm. Math. Phys. {\bf 187} (1997) 679, hep-th/9701162.

\item
\label{FMW99}
R. Friedman, J. Morgan and E. Witten,
{\em Vector bundles over elliptic fibrations}, 
J. Algebraic Geom., {\bf 8} (1999), pp.~279--401,  alg-geom/9709029.

\item
\label{DOPW1}
R. Donagi, B. Ovrut, T. Pantev and D. Waldram,
{\em Standard-Model Bundles on Non-Simply Connected Calabi--Yau Threefolds},
JHEP {\bf 0108} (2001) 053, hep-th/0008008.

\item
\label{DOPW2}
R. Donagi, B. Ovrut, T. Pantev and D. Waldram,
{\em Standard-model bundles},
Adv.Theor.Math.Phys. {\bf 5} (2002) 563, math.AG/0008010.

\item
\label{DOPW3}
R. Donagi, B. Ovrut, T. Pantev and D. Waldram,
{\em Spectral involutions on rational elliptic surfaces},
Adv.Theor.Math.Phys. {\bf 5} (2002) 499, math.AG/0008011.

\item
\label{DHOR}
R. Donagi, Y.-H. He, B.A. Ovrut and R. Reinbacher,
{\em The Spectra of Heterotic Standard Model Vacua},
JHEP {\bf 0506} (2005) 070, hep-th/0411156.

\item
\label{BD}
V. Bouchard and R. Donagi,
{\em An SU(5) Heterotic Standard Model},
Phys.Lett. {\bf B633} (2006) 783,
hep-th/0512149.

\item
\label{AH}
B. Andreas and D. Hern\'andez Ruip\'erez,
{\em U(n) Vector Bundles on Calabi-Yau 
Threefolds for String Theory Compactifications}, 
Adv. Theor. Math. Phys. {\bf 9} (2006) 253, hep-th/0410170.

\item
\label{BDO}
E.~Buchbinder, R.~Donagi, and B.~A. Ovrut, {\em  Vector bundle moduli and
small instanton transitions}, JHEP 0206 (2002) 054,
hep-th/0202084.

\item
\label{CD}
G. Curio and R. Donagi, Nucl. Phys. {\bf B518} (1998) 603, hep-th/ 9801057.

\item
\label{an}
B. Andreas, 
{\em On vector bundles and chiral matter in N=1 heterotic compactifications},
JHEP {\bf 9901} (1999) 011, hep-th/9802202.

\item
\label{ACK}
B. Andreas, G. Curio and A. Klemm,
{\em Towards the Standard Model from elliptic Calabi-Yau},
Int.J.Mod.Phys. {\bf A19} (2004) 1987, hep-th/9903052.

\item
\label{C c3}
G. Curio, {\em Chiral matter and transitions in heterotic string models},
hep-th/9802224, Phys.Lett. {\bf B435} (1998) 39.

\item
\label{C}
G. Curio,
{\em Standard Model bundles of the heterotic string},
hep-th/0412182, to appear in Int.J.Mod.Phys. {\bf A} (2005).

\item
\label{Distler}
J.Distler and B.R. Greene, {\em Aspects of (2,0) String
Compactifications},
Nucl. Phys {\bf B304} (1988) 1.

\item
\label{Dine}
M. Dine, N. Seiberg, X.G. Wen and E. Witten, {\em Nonperturbative
Effects on the string world-sheet,2},
Nucl. Phys {\bf B289} (1987) 319.

\item
\label{Lukas}
A. Lukas and K.S. Stelle,
{\em Heterotic anomaly cancellation in five dimensionas},
JHEP {\bf 01} (2000) 010, hep-th/9911156.

\item
\label{Erler}
J. Erler, {\em Anomaly Cancellation in Six Dimensions},
J.Math.Phys. {\bf 35} (1994) 1819, hep-th/9304104.

\item
\label{blume}
R. Blumenhagen, G. Honecker and T. Weigand, 
{\em Lopp-corrected compactifications of the 
heterotic string with line bundles}, JHEP {\bf 0506} (2005) 020,
hep-th/0504232.

\item
\label{Timo}
T. Weigand, {\em Heterotic Vacua from general (non-) Abelian Bundles},
hep-th/0512191.

\item
\label{Blumenhagen06}
R. Blumenhagen, S. Moster and T. Weigand, 
{\em Heterotic GUT and Standard Model Vacua 
from simply connected Calabi-Yau Manifolds},
hep-th/0603015.

\item
\label{UhYa}
K.~Uhlenbeck and S.-T. Yau, {\em On the existence of
Hermitian Yang-Mills connections in stable vector bundles}, Comm. Pure
Appl. Math., 39 (1986), pp.~S257--S293.
\newblock Frontiers of the mathematical sciences: 1985 (New York, 1985).

\item
\label{Donald}
Donaldson, Proc. London Math. Soc. 3, 1 (1985).

\item
\label{Bjoern}
B. Andreas and D. Hern\'andez Ruip\'erez,
{\em Comments on N=1 Heterotic String Vacua},
Adv.Theor.Math.Phys. {\bf 7} (2004) 751,
hep-th/0305123.

\end{enumerate}

\end{document}